\documentclass[aps,pra]{revtex4}
\usepackage{graphicx}

\begin{document}

\title{Mutual information, bit error rate  and  security in W\'{o}jcik's scheme
\thanks{E-mail: Zhangzj@wipm.ac.cn} }

\author{
Zhanjun Zhang  \\
{\footnotesize  Wuhan Institute of Physics and Mathematics, The
Chinese Academy of Sciences, Wuhan 430071, China \\
E-mail: Zhangzj@wipm.ac.cn }}

\date{\today}

\maketitle

\begin{minipage}{380pt}
In this paper the correct calculations of the mutual information
of the whole transmission, the quantum bit error rate (QBER) are
presented. Mistakes of the general conclusions relative to the
mutual information, the quantum bit error rate (QBER) and the
security in W\'{o}jcik's paper [Phys. Rev. Lett. {\bf 90},
157901(2003)] have been pointed out. \\

PACS Number(s): 03.67.Hk, 03.65.Ud\\
\end{minipage}

After the pioneering work of Bennett and Brassard published in
1984[1], a variety of quantum secret communication protocols have
been proposed( for a review see [2]). Recently, a quite different
quantum cryptographic protocol (namely, the 'ping-pong' protocol)
has been proposed by Bostr\"{o}m and Felbinger[3], which allows
the generation of a deterministic key or even direct secret
communication. The protocol has been claimed to be secure and
experimentally feasible. However, since the security of the
'ping-pong' protocol can be impaired as far as considerable
quantum losses are taken into account, very recently W\'{o}jcik
has presented an undetectable eavesdropping scheme on the
'ping-pong' protocol[4].  The aims of this paper are as follows:
(1) to present the correct calculations of various mutual
information (i.e., the mutual information between the legitimate
sender (Alice) and the legitimate receiver (Bob) and the mutual
information between Alice and the eavesdropper (Eve)) and the
quantum bit error rates (QBERs, i.e., Bob's BQER and Eve's QBER);
(2) to point out the mistakes relative to the mutual information,
the QBERs and the security in W\'{o}jcik's paper and correct them.

Since W\'{o}jcik's scheme is a realistic scheme,  all the total
numbers of Alice's bits, Bob's bits and Eve's bits should be
finite in his paper. This is important in pointing out the
mistakes in W\'{o}jcik's paper. This can be seen later. The mutual
information in W\'{o}jcik's paper is only and essentially a
single-bit mutual information. This physical quantity is used
inappropriately in W\'{o}jcik's paper to stand for the mutual
information of the whole transmissions (multi bits),
alternatively, the mutual information of the whole transmissions
(multi bits) are not worked out appropriately in W\'{o}jcik's
paper. This can also be seen later. So, in this paper, first, let
us give the formulae of the mutual information of the whole
transmissions and the QBERs. Incidentally, one should keep in mind
that, provided that Alice's bits are given, the mutual information
between Alice and Bob (Eve) and the QBER in Bob's (Eve's) bits
should be completely determined by the bits Bob (Eve) obtains.

Let $J$ be the total number of Alice's bits and $J_0$ be the
number of all '0' bits and $J_1$ be the number of all '1' bits in
Alice's bits. Obviously, $J=J_0+J_1$. Let $a_0 (a_1)$ be the rate
of the '0' ('1') bit in Alice's bits, then $a_0=J_0/J$ and
$a_1=J_1/J=1-a_0$. If an assumption that $a_0=a_1=1/2$ is
employed, then $J_0=J_1=J/2$. In this paper such an assumption is
employed hereafter.

Let $M$ be the total number of Bob's bits and $M_0$ be the number
of all '0' bits and $M_1$ be the number of all '1' bits in Bob's
bits. Obviously, $M=M_0+M_1$. Let $b_0 (b_1)$ be the rate of the
'0' ('1') bit in Bob's bits, then $b_0=M_0/M$ and
$b_1=M_1/M=1-b_0$.

Obviously, only if an ideal quantum channel is assumed, then
$M=J$;  otherwise, $M$ should be less than $J$ due to the channel
losses.  In this paper, later, $J$ is assumed to be the effective
number of Alice's bits which can be transmitted to Bob. In this
case $M=J$.

Let $L_{00}$ be the total number extracted from Alice's $J_0$ '0'
bits and Bob's $M_0$ '0' bits in the case that when Alice sends a
'0' bit Bob accordingly gets  a '0' bit by his measurement.
$L_{00}$ is named as the number of the pair (0,0). Similarly,
$L_{10}$, $L_{01}$ and $L_{11}$ can be defined. Their rates can be
worked out as follows:
\begin{eqnarray}
c_{00}=L_{00}/J, \ \ \ c_{01}=L_{01}/J, \ \ \ c_{10}=L_{10}/J,\ \
\ c_{11}=L_{11}/J.
\end{eqnarray}

According to above definitions, the following relations can be
built up easily:
\begin{eqnarray}
L_{00}+L_{01}=J_0=J/2,
\end{eqnarray}
\begin{eqnarray}
L_{10}+L_{11}=J_1=J/2,
\end{eqnarray}
\begin{eqnarray}
L_{00}+L_{10}=M_0,
\end{eqnarray}
\begin{eqnarray}
L_{01}+L_{11}=M_1=J-M_0.
\end{eqnarray}

Let $Q_b$ be the number of the wrong bits in Bob's bits comparing
with Alice's bits. Easily, one can obtain $Q_b=L_{01}+L_{10}$.
Accordingly, the QBER in Bob's bits can be obtained:
\begin{eqnarray}
q_b \equiv Q_b/J = (L_{01}+L_{10})/J.
\end{eqnarray}

Taking advantage of equations 1-6, the following rates can be
arrived at
\begin{eqnarray}
\left \{ \begin{array}{c}
c_{00}=(2b_0+1-2q_b)/4, \ \ \ c_{01}=(1-2b_0+2q_b)/4, \\
c_{10}=(2b_0-1+2q_b)/4, \ \ \ c_{11}=(3-2b_0-2q_b)/4.
\end{array} \right.
\end{eqnarray}
It seems that the values of $b_0$ and $q_b$ can be chosen freely
in the domain $[0,1]$, however, since $0 \leq
c_{00},c_{01},c_{10},c_{11},b_0,q \leq 1$, the following
limitations exist:
\begin{eqnarray}
\left \{ \begin{array}{c} \frac{1}{2}-q_b \leq b_0 \leq
\frac{1}{2}+q_b, \ \ \ \ \
0 \leq q_b \leq \frac{1}{2};  \\
q_b-\frac{1}{2} \leq b_0 \leq \frac{3}{2}-q_b ,\ \ \ \ \
\frac{1}{2} \leq q_b \leq 1.
\end{array} \right.
\end{eqnarray}

According to the definition of the mutual information, one can
obtain the mutual information between Alice and Bob as follows:
\begin{eqnarray}
I_{AB} &\equiv& H(A:B)=H(A)+H(B)-H(A,B) \nonumber \\
&=& -\frac{1}{2}\log_2\frac{1}{2}-\frac{1}{2}\log_2\frac{1}{2}-b_0\log_2b_0-(1-b_0)\log_2(1-b_0) \nonumber \\
&+&\frac{2b_0+1-2q_b}{4}\log_2\frac{2b_0+1-2q_b}{4}+\frac{1-2b_0+2q_b}{4}\log_2\frac{1-2b_0+2q_b}{4}  \nonumber \\
&+&\frac{2b_0-1+2q_b}{4}\log_2\frac{2b_0-1+2q_b}{4}+\frac{3-2b_0-2q_b}{4}\log_2\frac{3-2b_0-2q_b}{4},
\end{eqnarray}
which is a function of $b_0$ and $q_b$. Once Bob finishes his
measurements, then in his bits the $b_0$ is determined and
accordingly the $q_b$ is determined provided that Alice's bits are
given. Up to now, the above equation has established the relation
between the mutual information and the QBER.

Let us see the properties of the equation (9): (a) If $q_b=0$,
according to the equation (8) one can obtain $b_0=1/2$.
Substituting $q_b=0$ and $b_0=1/2$ into the equation (9), one can
obtain $I_{AB}=1$. This is easily understood. Since Bob gets the
whole bits correctly, it is sure for him to get the whole
information. (b) If $q_b=1/2$, then $I_{AB} \equiv 0$ disregarding
the value of $b_0$ completely. This also can be seen from figure
1. In fact, when Bob gets his bits simply by guess instead of his
measurements, it is possible for him to get $J/2$ wrong bits
(i.e., $q_b=1/2$). In this case, he should get no information.
Incidentally, although it is very possible for Bob to get $J/2$
wrong bits (i.e., $q_b=1/2$) by guess, there are other
possibilities. (c) Substituting $1-q_b$ for $q_b$ in the equation
(9), the equation (9) does not change. This means that if Bob can
get the mutual complementary bits of his bits he may have the same
mutual information. This can be easily seen from figure 1. The
figure is symmetric about the $q_b=1/2$. Specifically, in both
cases of $q_b=0$ and $q_b=1$, $I_{AB}=1$. Therefore, when one uses
the mutual information as the criteria to judge how much
information Bob has obtained, an assumption that the mutual
complementary bits should correspond to the same information
(e.g., if '00100' defines a character, then '11011' as the
former's mutual complementary bits should defines the same
character) is employed; otherwise, such a criteria is incorrect.
Since the assumption is disadvantageous for the message
transmission in reality, the mutual information is not the
quantity which is good enough to character the successfully
transmitted message. By the way, the strategy combining the mutual
information with the QBER is feasible as an improvement. In
addition, as mentioned before, once Bob gets more or less wrong
bits relative to $J/2$ by his guess, it is possible for him to get
some information responding to his naive guess. (d) For a given
$q_b$, $I_{AB}$ is a function of $b_0$, that is, it is possible
that two different $b_0$'s do not correspond to a same $I_{AB}$
but a same $q_b$ (e.g. see the lines 7 and 8 in table 1), the .
This means that the QBER is not a suitable quantity in
characterizing the amount of the information Bob gains form Alice.

Similarly, one can work out the mutual information between Alice
and Eve. Let $K$ be the total number of Eve's bits and $K_0$ be
the number of all '0' bits and $K_1$ be the number of all '1' bits
in Eve's bits. Obviously, $K=K_0+K_1$. Let $e_0 (e_1)$ be the rate
of the '0' ('1') bit in Eve's bits, then $e_0=K_0/K$ and
$e_1=K_1/K=1-e_0$.

Let $N_{00}$ be the total number extracted from Alice's $J_0$ '0'
bits and Eve's $K_0$ '0' bits in the case that when Alice sends a
'0' bit Eve accordingly gets  a '0' bit by her eavesdropping.
$N_{00}$ is named as the number of the eavesdropping pair (0,0).
Similarly, $N_{10}$, $N_{01}$ and $N_{11}$ can be defined. Their
rates can be worked out as follows:
\begin{eqnarray}
d_{00}=N_{00}/K, \ \ \ d_{01}=N_{01}/K, \ \ \ d_{10}=N_{10}/K,\ \
\ d_{11}=N_{11}/K.
\end{eqnarray}

Let $Q_e$ be the number of the wrong bits in Eve's bits comparing
with Alice's bits. Easily, one can obtain $Q_e=N_{01}+N_{10}$.
Accordingly, the QBER in Eve's bits can be obtained:
\begin{eqnarray}
q_e \equiv Q_e/J = (N_{01}+N_{10})/J.
\end{eqnarray}

Assume Eve can attack all the bits, then $K=J$. According to the
definition of the mutual information, one can obtain the mutual
information between Alice and Eve as follows:
\begin{eqnarray}
I_{AE} &\equiv& H(A:E)=H(A)+H(E)-H(A,E) \nonumber \\
&=& -\frac{1}{2}\log_2\frac{1}{2}-\frac{1}{2}\log_2\frac{1}{2}-e_0\log_2e_0-(1-e_0)\log_2(1-e_0) \nonumber \\
&+&\frac{2e_0+1-2q_e}{4}\log_2\frac{2e_0+1-2q_e}{4}+\frac{1-2e_0+2q_e}{4}\log_2\frac{1-2e_0+2q_e}{4}  \nonumber \\
&+&\frac{2e_0-1+2q_e}{4}\log_2\frac{2e_0-1+2q_e}{4}+\frac{3-2e_0-2q_e}{4}\log_2\frac{3-2e_0-2q_e}{4}.
\end{eqnarray}

Let us turn to point out the mistakes in W\'{o}jcik's paper.
First, see two obvious facts:

(i) in W\'{o}jcik's paper, Bob's QBER and Eve's QBER are precisely
1/4. Then one would like to ask how many wrong bits in Bob's
(Eve's) bits provided that the total bit number $M$ is odd or
$2I+2$. Assuming $M=201,202,203$, whether the corresponding
answers are 50.25,50.50,50.75 respectively? If so, how ridiculous
they are for the numbers of the wrong bits are noninteger. It is
the wrong QBER=1/4 that leads to the ridiculous results.

(ii) According to W\'{o}jcik's figure 4, one can find when the
channel transmission efficiency is zero, both Eve and Bob still
can get information from Alice and Eve's is larger than Bob's.
Another ridiculous result!

All these have obviously shown that there are really mistakes in
W\'{o}jcik's paper. In fact, there are more mistakes in his paper.
These can be seen as follows. In the ping-pong protocol with ideal
quantum channel, Bob's bits are obtained uniquely and
deterministically, which are same as Alice's. For example, if
Alice's bits are '100110', then Bob gets '100110' after his
measurements. In W\'{o}jcik eavesdropping scheme, due to Eve's
attacks with help of channel losses, Bob's (Eve's) bits are not
unique in theory, and only after his (her) measurements his (her)
bits are determined. For an example: Let 'u' ('s') be Eve's attack
without (with) the symmetry operation. Assume Alice's bits are
'100110' and Eve's attacks are 'susuus'. Also Assume that the
transmission efficiency $\eta$ of the quantum channel is not
greater than 50\%, for in this case Eve can attack also the bits.
According to W\'{o}jcik's scheme, taking advantage of the
following conditional probability distributions:
\begin{eqnarray}
p^u_{000}=1, \,\, p^u_{001}=p^u_{010}=p^u_{011}=0, \,\,\,\,\,\,
p^u_{100}=p^u_{101}=p^u_{110}=p^u_{111}=1/4; \nonumber \\
p^s_{000}=p^s_{001}=p^s_{010}=p^s_{011}=1/4, \,\,\,\,\,\,\,
p^s_{111}=1, \,\, p^s_{100}=p^s_{101}=p^s_{110}=0;
\end{eqnarray}
in theory it is possible for Eve to get any one of the following
batches: '100110', '100111', '100100', '100101', '100010',
'100011', '100000', '100001','101100', '101101', '101110',
'101111', '101000', '101001','101010' and '101011'. According to
equation (12), it is easy to work out the mutual information for
each batch of bits. Obviously these batches do not lead to a same
value of $I_{AE}$ and a same QBER (See table 1).  This example
denies W\'{o}jcik's conclusion that the $I_{AE}$ is always 0.311
and the QBER is always 1/4. It is very easy to be understood that
when Eve gets different batches of bits it is quite possible for
her to have different mutual information with Alice (See table 1).
However, which batch of bits Eve will get on earth by his
measurements can not be determined beforehand, for each one can
occur with its own probability (See table 1). Only after Eve's
measurements, one can know which one it is exactly. From table 1,
one can also deny W\'{o}jcik's another general conclusion that
$I_{AE}$ is always larger than $I_{AB}$, for in some cases
$I_{AE}=0$ (See Table 1). So there must be mistakes in
W\'{o}jcik's calculations of the mutual information. I think it is
the confusion between the single-bit mutual information and the
multi-bits mutual information which leads to the mistakes. Hence,
intuitively, the security estimation based on the wrong mutual
information in W\'{o}jcik's paper is also not reliable anymore.

Let $J_u (J_s)$ be the number of Alice's bits suffering the 'u'
('s') attacks, then $J=J_u+J_s \equiv
J_{u0}+J_{u1}+J_{s0}+J_{s1}$. Further, still assume that after
Eve's 'susuus' attacks, Eve gets her '100110' from Alice's
'100110'. Then $J_{u0}=1$, $J_{u1}=2$, $J_{s0}=2$,$J_{u1}=1$,
$N^u_{00}=1$,$N^u_{01}=0$,$N^u_{10}=0$,$N^u_{11}=2$,
$N^s_{00}=2$,$N^s_{01}=0$,$N^s_{10}=0$,$N^s_{11}=1$,. The
extracted probability distributions of the eavesdropping pairs
(0,0),(0,1),(1,0) and (1,1) can be calculated as follows:
\begin{eqnarray}
t^u_{00}\equiv \frac{N_{00}^u}{J_{u0}}=1, \,\, t^u_{01}\equiv
\frac{N_{01}^u}{J_{u0}}=0, \,\, t^u_{10}\equiv
\frac{N_{10}^u}{J_{u1}}= 0,\,\, t^u_{11}\equiv
\frac{N_{11}^u}{J_{u1}}=1; \nonumber \\
t^s_{00}\equiv \frac{N_{00}^s}{J_{s0}}=1, \,\, t^s_{01}\equiv
\frac{N_{01}^s}{J_{s0}}=0, \,\, t^s_{10}\equiv
\frac{N_{01}^s}{J_{s0}}= 0,\,\, t^s_{11}\equiv
\frac{N_{11}^s}{J_{s1}}=1.
\end{eqnarray}
They are apparently different with the conditional probability
distributions:
\begin{eqnarray}
t^u_{00}= p^u_{000}+ p^u_{001},\,\,p^u_{01}= p^u_{010}+
p^u_{011},\,\, t^u_{10} \neq p^u_{100}+ p^u_{101},\,\,
t^u_{11} \neq p^u_{110}+ p^u_{111}; \nonumber \\
t^s_{00} \neq p^s_{000}+ p^s_{001},\,\, p^s_{01}\neq p^s_{010}+
p^s_{011},\,\, t^s_{10} = p^s_{100}+ p^s_{101}, \,\, t^s_{11} =
p^s_{110}+ p^s_{111}.
\end{eqnarray}
The essential reason is that the number $J=6$ of the whole bits is
finite. By the way, since Wojick's scheme is a realistic scheme,
$J$ should be finite. Generally speaking, when $J$ is large
enough, then the extracted probability distributions may be close
to the conditional probability distributions. Only when all the
$J$, $J_{u1}$ and $J_{s0}$ are infinite, the extracted probability
distributions are equivalent to the conditional probability
distributions, that is,
\begin{eqnarray}
t^u_{00}\equiv \frac{N_{00}^u}{J_{u0}}= p^u_{000}+ p^u_{001}=1,
\,\, t^u_{01}\equiv \frac{N_{01}^u}{J_{u0}}=p^u_{010}+
p^u_{011}=0, \nonumber \\
t^u_{10}\equiv \frac{N_{10}^u}{J_{u1}}= p^u_{100}+
p^u_{101}=\frac{1}{2},\,\, \,\,t^u_{11}\equiv
\frac{N_{11}^u}{J_{u1}}=p^u_{110}+ p^u_{111}=\frac{1}{2}; \nonumber \\
t^s_{00}\equiv \frac{N_{00}^s}{J_{s0}}=p^s_{000}+
p^s_{001}=\frac{1}{2}, \,\, t^s_{01}\equiv
\frac{N_{01}^s}{J_{s0}}=p^s_{010}+ p^s_{011}=\frac{1}{2}, \nonumber \\
t^s_{10}\equiv \frac{N_{10}^s}{J_{s0}}= p^s_{100}+
p^s_{101}=0,\,\, \,\,t^s_{11}\equiv
\frac{N_{11}^s}{J_{s1}}=p^s_{110}+ p^s_{111}=1.
\end{eqnarray}
Assume that $J_{u0}=J_{u1}=J_{s0}=J_{s1}=J/4$ (same hereafter),
then one can obtain
\begin{eqnarray}
e_0=K_0/J=(N_{00}+N_{10})/J=(N^u_{00}+N^u_{10}+N^s_{00}+N^s_{10})/J=1/2,
\nonumber \\
q_e=Q_e/J=(Q_{eu}+Q_{es})/J=(N^u_{01}+N^u_{10}+N^s_{01}+N^s_{10})/J=1/4.
\end{eqnarray}
Substituting $e_0=1/2$ and $q_e=1/4$ into the equation (12), then
one can obtain $I_{AE}=\frac{3}{4}\log_23-1\approx 0.189$, which
is different from the value W\'{o}jcik claims. This is easily
understood. As I have mentioned before, in W\'{o}jcik's paper, the
mutual information are essentially single-bit mutual information.
Since Eve knows exactly when  the 's' operations are performed,
for each bit the mutual information can be easily worked out and
really not change. However, such single-bit mutual information is
not suitable for representing the multi-bit mutual information.
From my deduction, one can see it is the inappropriate use of the
mutual information which leads to the wrong result in W\'{o}jcik's
paper. Accordingly, there are mistakes in W\'{o}jcik's figure 4.
Intuitively, the security estimation based on the wrong mutual
information in W\'{o}jcik's paper is also not reliable anymore.

To summarize, I have presented the calculations of the multi-bit
mutual information and the QBERs. A number of mistakes in
W\'{o}jcik's have been found and pointed out.

This work is supported by the National Natural Science Foundation
of China under Grant No. 10304022.

\begin{center}

\noindent Table 1 The possibility, the QBER, the rate of '0' bit in Eve's bits
and the mutual information between Alice and Eve for possible batches of bits. \\
\begin{tabular}{ccccccc} \hline
Eve's bits & possibility & $q$ && $e_0$ && $I_{AB}$ \\ \hline
'100110'   & 1/16        &  0  &&  1/2  &&    1     \\
'100111'   & 1/16        & 1/6 &&  1/3  &&   0.459  \\
'100100'   & 1/16        & 1/6 &&  2/3  &&   0.459  \\
'100101'   & 1/16        & 1/3 &&  1/2  &&   0.082  \\
'100010'   & 1/16        & 1/6 &&  2/3  &&   0.459  \\
'100011'   & 1/16        & 1/3 &&  1/2  &&   0.082  \\
'100000'   & 1/16        & 1/3 &&  5/6  &&   0.134  \\
'100001'   & 1/16        & 1/3 &&  2/3  &&   0.093  \\
'101100'   & 1/16        & 1/3 &&  1/2  &&   0.082  \\
'101101'   & 1/16        & 1/2 &&  1/3  &&     0    \\
'101110'   & 1/16        & 1/6 &&  1/3  &&   0.459  \\
'101111'   & 1/16        & 1/3 &&  1/6  &&   0.093  \\
'101000'   & 1/16        & 1/2 &&  2/3  &&     0    \\
'101001'   & 1/16        & 2/3 &&  1/2  &&   0.082  \\
'101010'   & 1/16        & 1/3 &&  1/2  &&  0.082   \\
'101011'   & 1/16        & 1/2 &&  1/3  &&    0     \\ \hline
\end{tabular} \\
Alice's bits  are '100110'. Eve's attacks are 'susuus'.  \\ The
transmission efficiency $\eta$ of the quantum channel is assumed
to be not greater than 50\%.
\end{center}
\vskip 2cm

Caption

Figure 1 \,\, The mutual information between Alice and Bob as a
function of $q$ and $b_0$. See text for $q$ and $b_0$. (a) surface
diagram; (b) contour diagram.

\begin{figure}
 \begin{center}
 \includegraphics[width=1.0\textwidth]{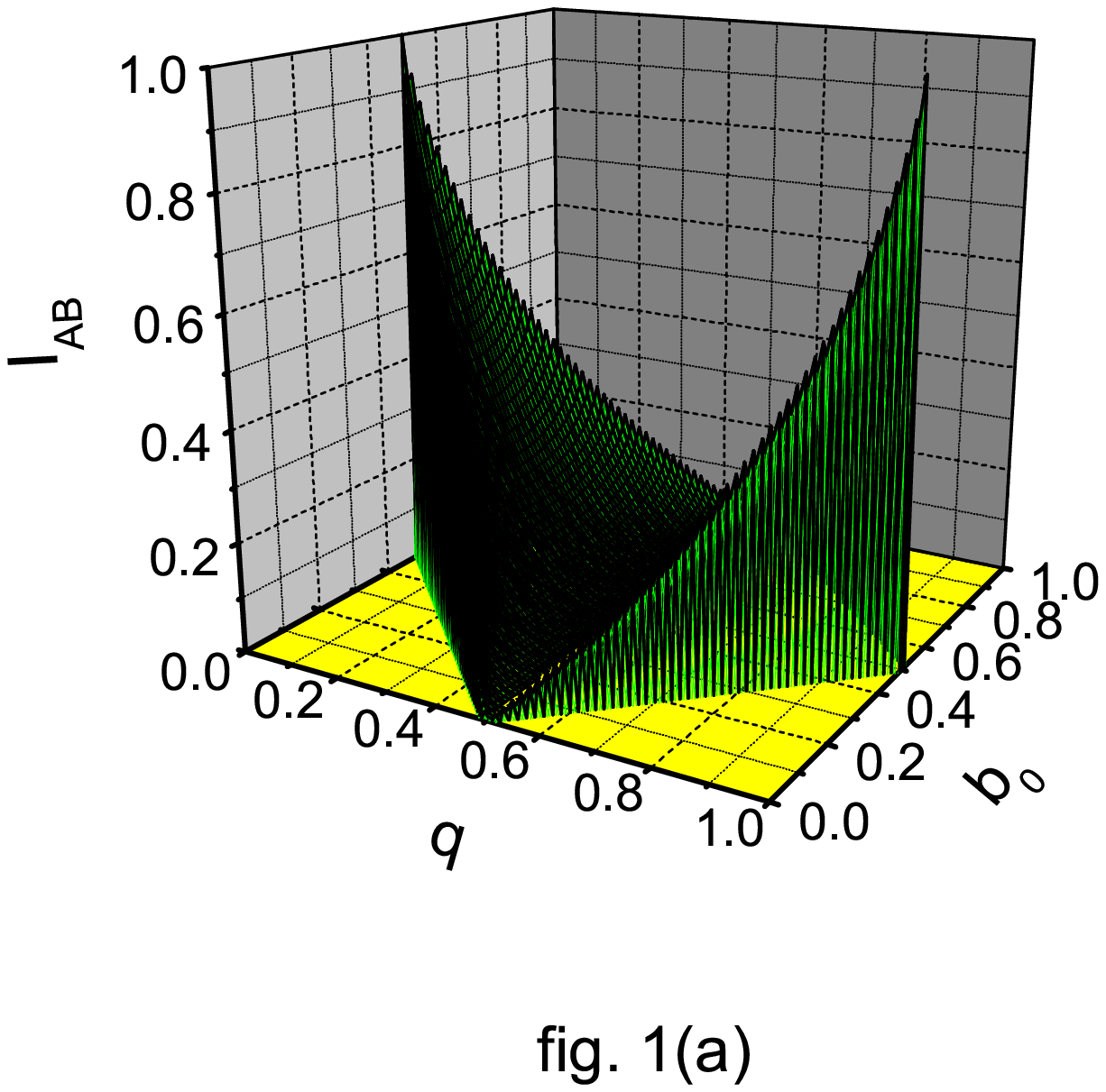}
 \caption{ }
 \label{}
 \end{center}
\end{figure}
\newpage

\begin{figure}
 \begin{center}
 \includegraphics[width=1.0\textwidth]{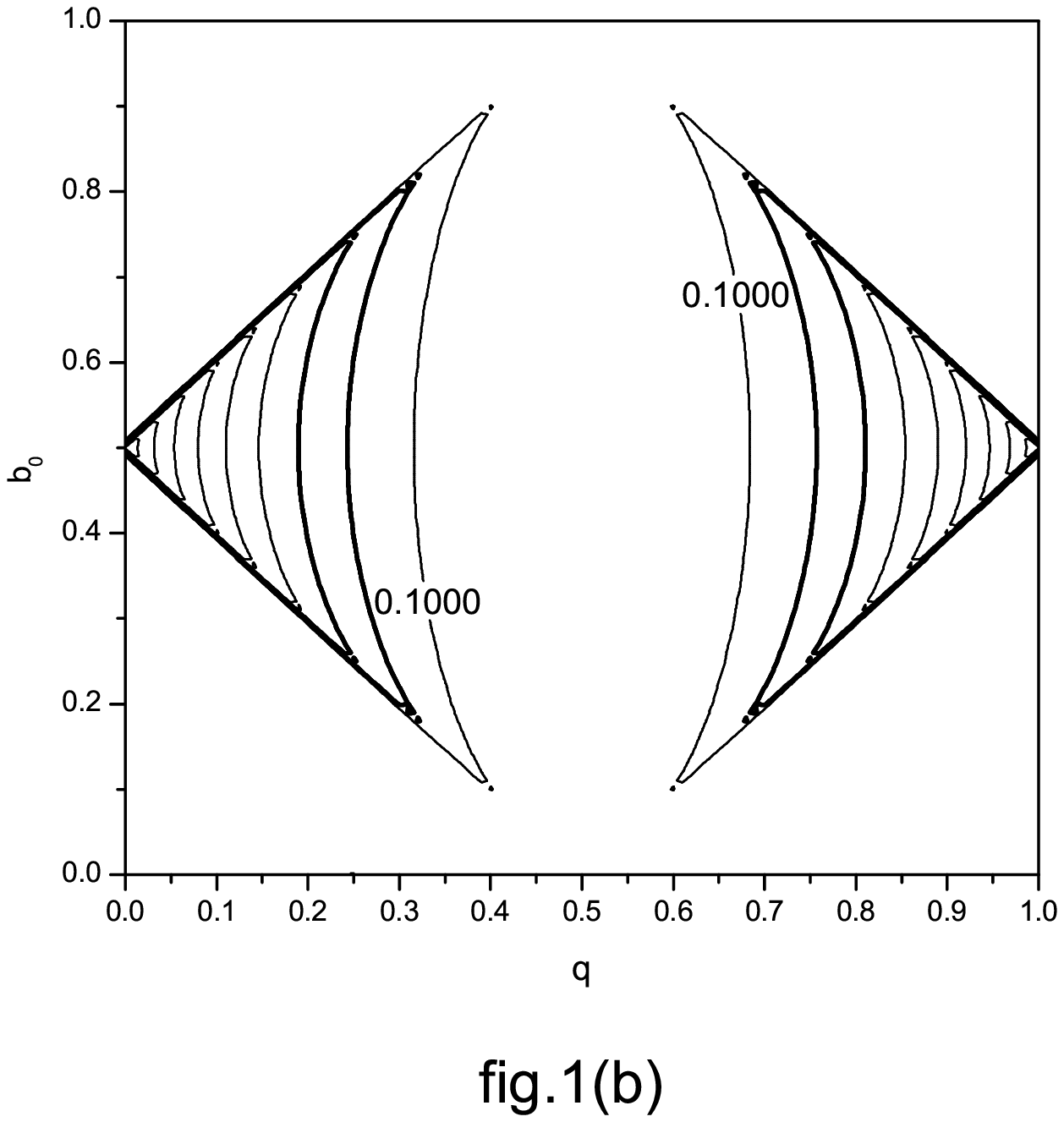}
 \caption{ }
 \label{}
 \end{center}
\end{figure}
\newpage
\end{document}